\documentclass[11pt]{article}
\usepackage{amssymb}
\usepackage{amsmath}
\usepackage{graphicx}
\newcommand{\be}{\begin{equation}}
\newcommand{\ee}{\end{equation}}
\newcommand{\bea}{\begin{eqnarray}}
\newcommand{\eea}{\end{eqnarray}}

\begin{document}

\title{\textbf{Stable Exact Cosmological Solutions in Induced Gravity Models}}

\author{Ekaterina~O.~Pozdeeva$^1$\footnote{pozdeeva@www-hep.sinp.msu.ru} \ and \    Sergey~Yu.~Vernov$^3$\footnote{svernov@theory.sinp.msu.ru}\\[2.7mm]
\small{${}^1$Skobeltsyn Institute of Nuclear Physics, Lomonosov  Moscow State University,}\\
\small{Leninskie Gory 1, 119991, Moscow, Russia}
}
\date { \ }
\maketitle

\begin{abstract}
We study dynamics of induced gravity cosmological models with sixth degree potential, that
have found using the superpotential method.
The important property of these models are existence of exact cosmological solutions
that tend to fixed points. The stability of these cosmological solutions have been obtained.
In particular, we find conditions under which solutions with a non-monotonic Hubble parameter that tend to a fixed point
are attractors.
\end{abstract}

\maketitle

\section{Introduction}

A large number of observational data~\cite{Perlmutter:1998np,Riess:1998cb,WMAP,Tegmark,Wood-Vasey,Jain:2003tba,Bernui:2005pz,
Planck2013} supports that the post-inflationary Universe is nearly homogeneous and isotropic and that the expansion of the Universe is accelerating.
It was also shown that the Universe is spatially flat at large distances.

There are two possible ways of describing the accelerating expansion of the Universe. The assumption that General Relativity is the
correct theory of gravity leads to the conclusion that there exists and currently dominates a smoothly distributed, slowly varying cosmic fluid with negative pressure,  so-called dark energy~\cite{DE_rev,DINDE,Wetterich:2014bma}.
The simplest way to describe  the dark energy is to add the cosmological constant to the  Einstein--Hilbert action.
Another popular variant is to consider models with scalar fields~\cite{DINDE,Wetterich:2014bma,Tsujikawa:2013fta}. There are a few reason to consider cosmological models with scalar fields. First of all there is no doubt that the evolution of the Universe  can be described by the spatially flat Friedmann--Lema\^{i}tre--Robertson--Walker (FLRW) background and cosmological perturbations. Models with scalar fields are very useful to describe such a type of the evolution. That is why scalar fields play an essential role in modern cosmology, in particular, in the current description of the evolution of the Universe at the early  epoch~\cite{Starobinsky:1979ty,Mukhanov:1981xt,Guth:1980zm,Linde:1981mu,Albrecht:1982wi,inflation2}.

To specify the cosmic fluid one typically uses a phenomenological relation $p=w\varrho$ between the
pressure (Lagrangian density) $p$ and the energy density $\varrho$
of this fluid. If a model includes only one cosmic fluid then in the FLRW metric the equation of state parameter $w$ and the Hubble parameter $H$ are connected as follows:
\begin{equation*}
 w={}-1-\frac{2\dot H}{3H^2},
\end{equation*}
where differentiation with respect to the cosmic time $t$ is denoted by a dot.

The cosmological constant corresponds to a constant $H$ (de Sitter Universe).
The standard way to obtain an evolving Hubble parameter is to add scalar
fields to a cosmological model. It has been demonstrated that dark energy with an evolving equation of state parameter provides a compelling alternative to a cosmological constant if data are analysed in a prior-free manner and the weak energy condition is not imposed by hand~\cite{9}.
For the model with minimally coupled scalar with the standard kinetic term we get $w>-1$ and the Hubble parameter
is a monotonically decreasing function. To get non-monotonic behaviour of the Hubble parameter
one needs either to add a phantom scalar field and get a quintom model~\cite{Guo2004,AKV2,Lazkoz,Quinmodrev1,Quinmodrev2}, or consider a scalar field with the second-derivative Lagrangian (see~\cite{Rubakov2014} as a review). Note that the dark energy
equation of state parameter can be less than $-1$, since this possibility is not excluded
by astrophysical data~\cite{9,Nesseris}. At the same time the cosmological models with $w<-1$ violate the null energy
condition (NEC), this violation is generally related to the phantom
fields appearing. The standard quantization of these models leads to
instability, which is physically unacceptable. Examples of cosmological models with nonstandard scalar fields that
admit solutions violating the NEC and having no obvious pathologies are presented~in~\cite{Rubakov2014}.

Note that consideration of models with phantom scalar fields can be motivated by
the string field theory~\cite{IA,AKV,McInnes,AKV2}.
The theory with $w<-1$ can be interpreted as an
approximation in the framework of the fundamental theory. Because the
fundamental theory must be stable and must admit quantization, this
instability can be considered an artefact of the approximation.
In the FLRW metric the NEC violating models can have classically stable solutions cosmology. In particular,
there are classically stable solutions for self-interacting ghost
models with minimal coupling to  gravity. Moreover, there exists an
attractor behavior in a class of the phantom cosmological
models~\cite{phantom-attractor,AKVCDM,ABJV09}.

Alternative way to describe the dark energy is to modify the theory of gravity~\cite{Fujii_Maeda,NO-rev,Book-Capozziello-Faraoni,CL}. Some of popular modified gravity models, for example, ${\cal F}(R)$ gravity models, can then be mapped into general relativity models with additional scalar fields by a suitable conformal transformation of the metric (see, e.g.,~\cite{Book-Capozziello-Faraoni,CL,Felice_Tsujikawa}). Another example is a class of nonlocally modified gravity models which were proposed to explain the current phase of cosmic acceleration without dark energy~\cite{Deser:2007jk} (see also~\cite{DW2009,DodelsonPark,Deser:2013uya,Foffa:2013vma} and the review~\cite{Woodard2014}). These models do not assume the existence of a new dimensional parameter in the action and include a function $f(\Box^{-1} R)$, with $\Box$ the d'Alembertian operator. These nonlocal models have a local scalar-tensor formulation that has been proposed in~\cite{Odintsov0708}, developed and used in~\cite{Jhingan:2008ym,Koivisto:2008,Non-local-FR,Sasaki,EPV}.

Predictions of  simplest inflationary models are  in
disagreement with the Planck2013 results~\cite{Planck2013}. At the same time many of these inflationary scenarios can be improved by adding a tiny non-minimal coupling of the inflaton field to gravity~\cite{GB2013,KL2013}. In the last years, the Higgs-driven inflation that includes non-minimal coupling between the Higgs boson and gravity has attracted a lot of attention~\cite{HI,Bezrukov:2008ut,HI1,HI3,HI4,Bezrukov2013}.

The models with the Ricci scalar multiplied by a function of the scalar field are being intensively studied in cosmology~\cite{Cooper:1982du,Kaiser,KKhT,Elizalde,Cerioni,Szydlo,Kamenshchik:2012rs,CervantesCota:2010cb,KTV2011,Sami:2012uh,ABGV,KTVV2013,
KPTVV2013} (see also~\cite{Fujii_Maeda,Book-Capozziello-Faraoni,CL} and references therein).
Generally these models are described by the following action:
\begin{equation}
\label{action}
S=\int d^4 x \sqrt{-g}\left[ U(\phi)R-\frac12g^{\mu\nu}\phi_{,\mu}\phi_{,\nu}-V(\phi)\right],
\end{equation}
where $U(\phi)$ and $V(\phi)$ are differentiable functions of the scalar field $\phi$, $g$ is the determinant of the metric tensor
$g_{\mu\nu}$ and the signature $(-,+,+,+)$ is used.

The reconstruction procedure for the induced gravity models ($U(\phi)=\xi\phi^2$, where $\xi$ is a constant) has been proposed in~\cite{KTV2011}.
In such a case it has been shown that one can linearize all the differential equations that should be solved in the reconstruction procedure to get the potential corresponding to a given cosmological evolution. This property allows one to obtain the explicit forms of potentials reproducing the dynamics of a spatially flat FLRW universe, driven by barotropic perfect fluids, by a Chaplygin gas  and by a modified Chaplygin gas~\cite{KTV2011}.

In~\cite{KTVV2013} another reconstruction procedure for the models described by  action~(\ref{action}) has been considered.
Such a method is similar to the Hamilton--Jacobi method (also known as the superpotential method or the first-order formalism) and is applied to cosmological models with scalar fields~\cite{Superpotential,AKV,Bazeia,AKV2,Andrianov:2007ua,ABV,Rotova,Harko:2013gha}. Note that similar method is used for the reconstruction procedure in brane models~\cite{DEWolfe,MMSV} and in holographic models
\cite{Gursoy:2008za,Aref'eva:2014sua}.
The key point in this method is that the Hubble parameter is considered as a function of the scalar field $\phi$. These two methods described above supplement each other and together allow one to construct different cosmological models with some required properties.

In this paper we study the stability of solutions in the FLRW metric, considering isotropic and homogenous fluctuations only.
In other word, we assume that not only background solution, but also perturbations are homogeneous and isotropic.
We analyse the stability with respect to small fluctuations of the initial data.

In~\cite{KTVV2013}, induced gravity models with the sixth degree polynomial potentials and physically interesting behaviors of the Hubble parameter have been obtained. To get these models the superpotential method has been used, because  the explicit form of the Hubble parameter $H(t)$ is too complicated to be guessed.
In particular, solutions with non-monotonic behaviors of the Hubble parameter have been obtained. In this paper we consider the stability of these solutions.

The stability of a continuous solution tending to a fixed point (kink or lump like solutions) implies
the stability of this fixed point. This fixed point is a de Sitter solution if the corresponding Hubble parameter is not equal to zero.
De Sitter solutions which are attractors of the system describing the cosmological dynamics of non-minimally coupled scalar field have been
investigated in~\cite{Sami:2012uh}. In this paper we use the Lyapunov theorem~\cite{Lyapunov,Pontryagin} and find
conditions under which the fixed point and the corresponding kink (or
lump) solution are stable. In these cases the necessary condition of  exact solution's stability is boundedness of the first corrections  for positive time semiaxis. In particular we find conditions under which solutions this nonmonotonic behaviour of the Hubble parameter are stable.

The paper is organized as follows.
In Section~2, we review the basic equations for a induced gravity model and remind the algorithm of the superpotential reconstruction procedure. In Section~3 we consider induced gravity cosmological models. In Section~4, we find solutions and potentials with a Hubble parameter that evolves toward a constant value at late times. In Section~5 we obtain the stability conditions.
In Section~6, we analyse the stability of the kink-type solutions.
Finally Section~7 is devoted to the conclusions.

\section{Cosmological models with non-minimally coupled scalar fields}

Let us consider the spatially flat FLRW universe with the interval
\begin{equation*}
ds^2={}-dt^2+a^2(t)\left(dx_1^2+dx_2^2+dx_3^2\right).
\end{equation*}
The  Friedmann equations derived by variation of action (\ref{action}) have the following form~\cite{KTV2011,KTVV2013}:
\begin{equation}
\label{Fr1}
6UH^2+6\dot U H=\frac{1}{2}\dot\phi^2+V,
\end{equation}
\begin{equation}
\label{Fr2}
2U\left(2\dot H+3H^2\right)+4\dot U H+2\ddot U +\frac{1}{2}\dot\phi^2-V=0,
\end{equation}
where the Hubble parameter is  the logarithmic derivative of the scale factor:
$H=\dot a/a$.
The variation of action (\ref{action}) with respect to $\phi$ gives
\begin{equation}
\label{Fieldequ}
\ddot \phi+3H\dot\phi+V^{\prime}=6\left(\dot H +2H^2\right)U^{\prime}\,,
\end{equation}
where the prime indicates the derivative with respect to the scalar field
$\phi$.
 Combining  Eqs. (\ref{Fr1}) and (\ref{Fr2}) we obtain
\begin{equation}
\label{Fr21}
4U\dot H-2\dot U H+2\ddot U +\dot\phi^2=0.
\end{equation}
From Eqs.~(\ref{Fr1})--(\ref{Fr21}), one can get the following system of the first order equations~\cite{ABGV}:
\begin{equation}
\begin{split}
  \dot\phi&=\psi,\\
\dot\psi&={}-3H\psi-\frac{\left[(6 U''+1)\psi^2-4V\right]U'+2UV'}{2\left(3 {U'}^2+ U\right)},\\
\dot H&={}-\frac{2U''+1}{4\left(3{U'}^2+U\right)}\psi^2+\frac{2U'}{{3{U'}^2+U}}H\psi
-\frac{6{U'}^2}{3{U'}^2+U}H^2+\frac{U'V'}{2\left(3{U'}^2+U\right)}\,.
\end{split}
\label{FOSEQU}
\end{equation}

Note that Eq.~(\ref{Fr1}) is not a consequence of system  (\ref{FOSEQU}). On the other hand,
if Eq.~(\ref{Fr1}) is satisfied in the initial moment of time, then from system (\ref{FOSEQU}) it follows
that Eq.~(\ref{Fr1}) is satisfied at any moment of time. In other words, system (\ref{FOSEQU}) is equivalent to the initial system of equations (\ref{Fr1})--(\ref{Fieldequ}) if and only if we choose such initial data that Eq.~(\ref{Fr1}) is satisfied.

Let $H=Y(\phi)$ and the function ${\cal F}(\phi)$ is defined as follows
\begin{equation}
\label{equsigma}
\dot \phi={\cal F}(\phi).
\end{equation}
Substituting $\dot\phi$ and $\ddot \phi={\cal F}^{\prime}{\cal F}$ into Eq.~(\ref{Fr21}), one obtains the following equation~\cite{KTVV2013}:
\begin{equation}
\label{equa}
4UY^{\prime}+2({\cal F}^{\prime}-Y)U^{\prime}+\left(2U^{\prime\prime}+1\right){\cal F}=0.
\end{equation}
Equation (\ref{equa}) contains three functions. If two of them are given,
then the third one can be found as the solution of a linear differential
equation.  Let us note that Eq.~(\ref{equa}) is a first order
differential equation for both  $Y$ and $F$.

The potential $V(\phi)$ can be obtained from (\ref{Fr1}):
\begin{equation}
\label{potentialV}
V(\phi)=6UY^2+6U^{\prime}{\cal F}Y-\frac{1}{2}{\cal F}^2.
\end{equation}
To find the function $\phi(t)$ and, hence, $H(t)=Y(\phi(t))$ we integrate Eq.~(\ref{equsigma}).

Let system (\ref{FOSEQU}) has a constant solution (fixed point): $(\phi_f,\psi_f,H_f)$.
For  such a solution $\psi_f=0$ and
\begin{equation}
\label{FP-cond}
    V(\phi_f)=6U(\phi_f)H_f^2, \qquad V'(\phi_f)=2\frac{U'(\phi_f)}{U(\phi_f)}V(\phi_f).
\end{equation}
In terms of functions ${\cal F}$ and $Y$ we get the following  conditions for fixed point:
\begin{equation}
\label{FP_phi}
    {\cal F}(\phi_f)=0, \qquad V(\phi_f)=6U(\phi_f)Y(\phi_f)^2, \qquad V'(\phi_f)=12U'(\phi_f)Y(\phi_f)^2.
\end{equation}

\section{Induced gravity cosmological models}

In this paper, we are interested in the induced gravity models~\cite{Kaiser,KTV2011} with
\begin{equation}
U(\phi)= \frac \xi 2 \phi^2\,,
\end{equation}
where $\xi$ is the non-minimal coupling constant.

Equations (\ref{Fr1})--(\ref{Fieldequ}) for such choice of the function $U(\phi)$ look as follows:
\begin{equation}
\label{e2}
H^2=\frac{V}{3\xi\phi^2}+\frac{1}{6\xi}\left(\frac{\dot{\phi}}{\phi}\right)^2-2H\frac{\dot{\phi}}{\phi},
\end{equation}
\begin{equation}
\label{Equ11}
3H^2+2\dot{H}={}-2 \frac{\ddot{\phi}}{\phi}-4H\frac{\dot{\phi}}{\phi}
-\frac{4\xi+1}{2\xi}\left(\frac{\dot{\phi}}{\phi}\right)^2+ \frac{V}{\xi\phi^2},
\end{equation}
\begin{equation}
\label{Equ_phi}
\ddot{\phi}+3H\dot{\phi}+V'-6\xi\phi\left(2H^2+\dot{H}\right)=0,
\end{equation}
and system~(\ref{FOSEQU}) has the following form:
\begin{equation}
\label{IG_SYSTEM}
\begin{split}
  \dot\phi&=\psi,\\
\dot\psi&={}-3H\psi-\frac{\psi^2}{\phi}+\frac{1}{(1+6\xi)\phi}[4V(\phi)-\phi V'(\phi)],\\
\dot{H}&=\frac{4H\psi}{(1+6\xi)\phi} +\frac{V'(\phi)}{(1+6\xi)\phi}-\frac{12\xi}{1+6\xi}H^2-\frac{1+2\xi}{2\xi(1+6\xi)}\left(\frac{\psi}{\phi}\right)^2. \end{split}
\end{equation}

It has been shown in~\cite{ABGV} that at $\xi= -1/6$ nontrivial solutions in the FLRW metric exist for the potential $V=V_0\phi^4$ only. By this reason, we assume that $\xi\neq -1/6$.

For induced gravity models we get from equation~(\ref{equa}) that
\begin{equation}\label{Fphi}
    {\cal F}(\phi) = \phi^{-1-1/(2\xi)}\left[B_1-\int\phi^{(2\xi+1)/(2\xi)}\left(Y'\phi-Y\right)d\!\phi\right],
\end{equation}
where $B_1$ is an arbitrary constant.

\section{Different behavior of the Hubble Parameter}

Let $Y(\phi)$ is a generic quadratic polynomial
\begin{equation}
\label{Y2}
Y(\phi)=C_0+C_1\phi+ C_2\phi^2,
\end{equation}
where $C_0$, $C_1$, and $C_2$ are arbitrary constants, but $C_2\neq 0$. From (\ref{Fphi}) we obtain
\begin{equation}
{\cal F}(\phi)=\frac{2\left((8\xi+1)C_0-(4\xi+1)C_2\phi^2\right)\xi\phi}{(4\xi+1)(8\xi+1)}+ B\phi^{-(1+2\xi)/(2\xi)},
\end{equation}
where $B$ is an arbitrary constant.
Note that the function ${\cal F}(\phi)$ does not depend on $C_1$. We assume that $\xi\neq -1/4$ and
$\xi\neq -1/8$. When $B=0$, the function ${\cal F}(\phi)$ is a cubic polynomial and  the general solution for Eq.~(\ref{equsigma}) can be written in terms of elementary functions.

For $C_0\neq 0$ we obtain
\begin{equation}\label{stt}
\phi_\pm(t)=\pm\frac{\sqrt{(8\xi+1)C_0}}{\sqrt{(8\xi+1)C_0e^{-\omega (t-t_0)}+(4\xi+1)C_2}}\,,
\end{equation}
where $\omega=4\xi C_0/(4\xi+1)$, $t_0$ is an arbitrary integration constant.

At $C_0=0$ the general solution for Eq.~(\ref{equsigma}) is
\begin{equation}
\label{phiA00}
\tilde{\phi}_\pm(t)=\pm\frac{\sqrt{8\xi+1}}{\sqrt{4\xi C_2(t-t_0)}}.
\end{equation}

Solutions that correspond to ${\cal F}(\phi)=0$ are
\begin{equation}
\label{phideSi}
\phi_{f_0}=0, \qquad
\phi_{f_\pm}=\pm\frac{\sqrt{(8\xi+1)C_0}}{\sqrt{(4\xi+1)C_2}},
\end{equation}
Note that $\phi_{f_0}$ is a singular point for system (\ref{IG_SYSTEM}).
For de Sitter solutions with $\phi_{f_\pm}$ the value of the Hubble parameter is given by (\ref{Y2}).
The corresponding potential is the sixth degree polynomial:
\begin{equation}
\label{V6}
\begin{split}
V(\phi)&=\frac{(16\xi+3)(6\xi+1)\xi}{(8\xi+1)^2}C_2^2\phi^6+\frac{6(6\xi+1)\xi}{8\xi+1}C_1C_2\phi^5
+\left[3\xi C_1^2+\frac{2(6\xi+1)(20\xi+3)\xi}{(8\xi+1)(4\xi+1)}C_0C_2\right]\phi^4+{}\\[2.7mm]
&
{}+\frac{6(6\xi+1)\xi}{4\xi+1}C_0C_1\phi^3+\frac{(16\xi+3)(6\xi+1)\xi}{(4\xi+1)^2}C_0^2\phi^2\,.
\end{split}
\end{equation}

\begin{figure}[h]
\centering
\includegraphics[width=72mm]{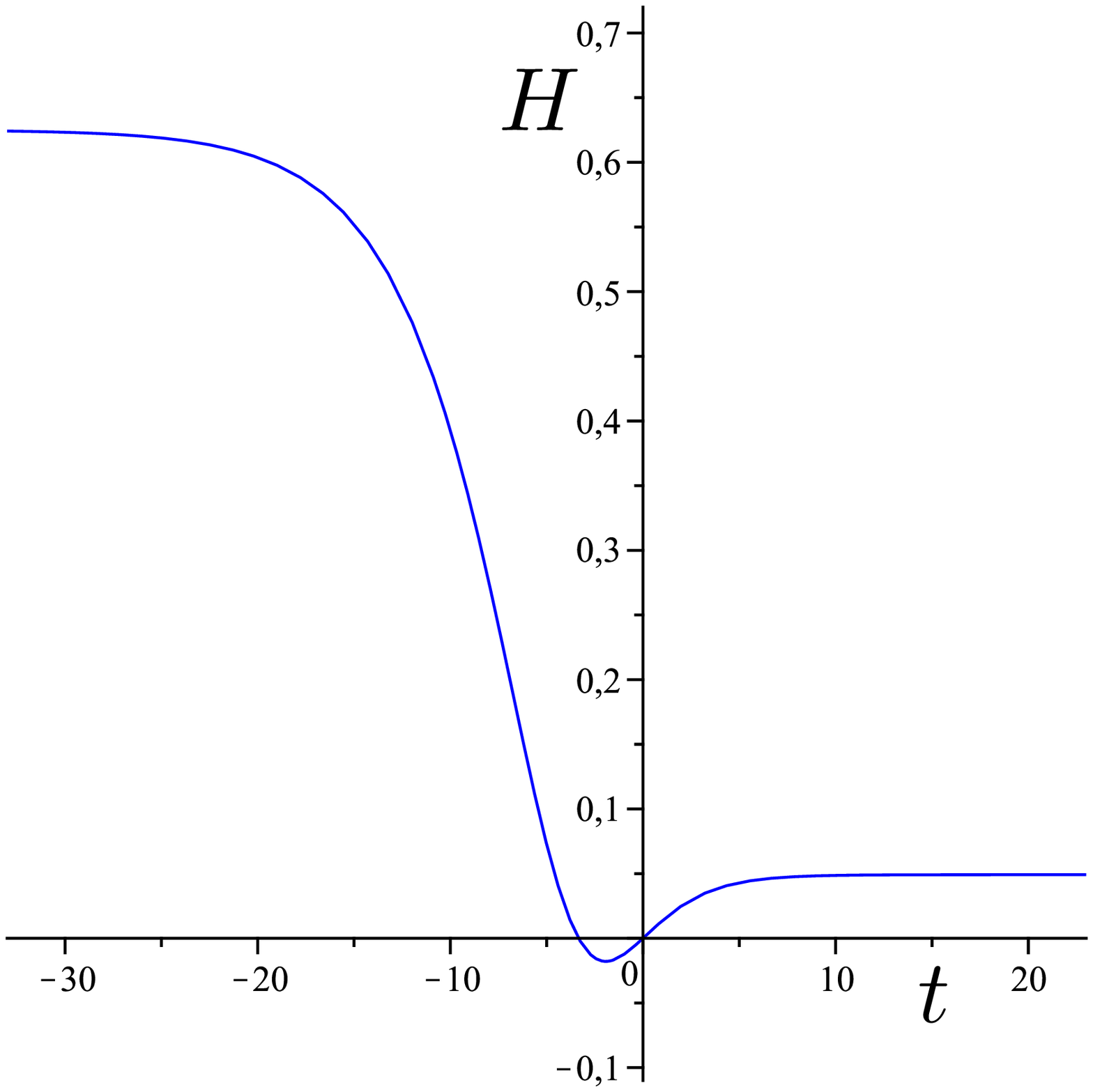} \ \ \ \ \ \ \
\includegraphics[width=72mm]{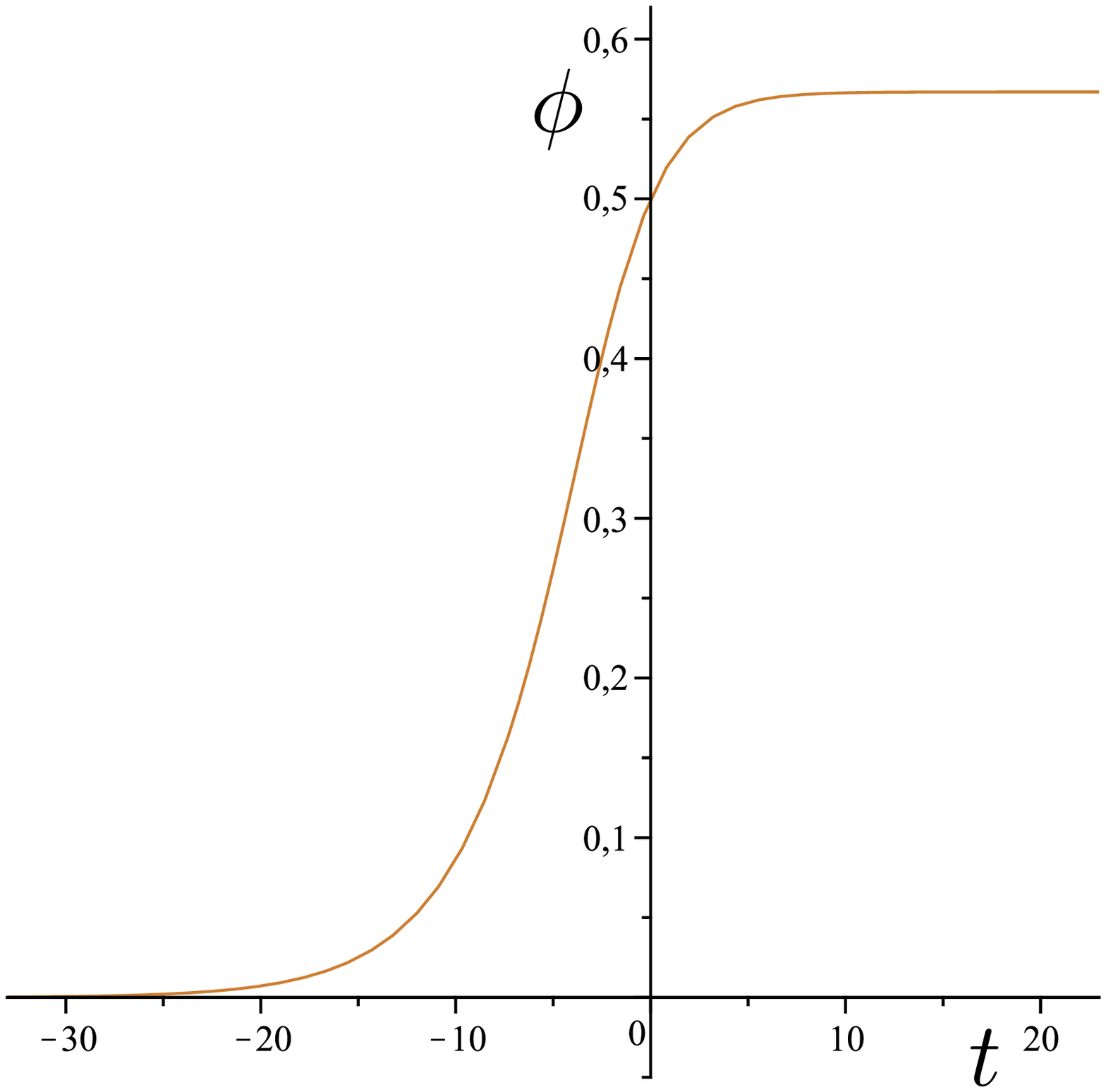}
\caption{The functions $H(t)$, given by (\ref{Y2}), and $\phi(t)=\phi_+(t)$ that correspond to a bounce solution.
The values of parameter are $\xi=1$, $C_2=7/2$, $C_1=-3$, and $C_0=5/8$, and $t_0=2\ln(8/9)\simeq -0.235$.   } \label{Fbounce}
\end{figure}

It is easy to find that there exist other fixed points with
\begin{equation*}
\tilde{\phi}_{f}=-\frac{1}{2C_2(4\xi+1)(16\xi+3)}\left[3C_1(8\xi+1)(4\xi+1)
\pm\sqrt{9C_1^2(8\xi+1)^2(4\xi+1)^2-4C_0C_2(16\xi+3)^2}\right]\!.
\end{equation*}
We do not know solutions that tend to these fixed points, by this reason we do not consider the stability of these points in our paper.
In Fig.~\ref{Fbounce} we demonstrate that this model has bounce solution: $H(0)=0$, $\dot H(0)>0$.
\begin{figure}[h]
\centering
\includegraphics[width=45mm]{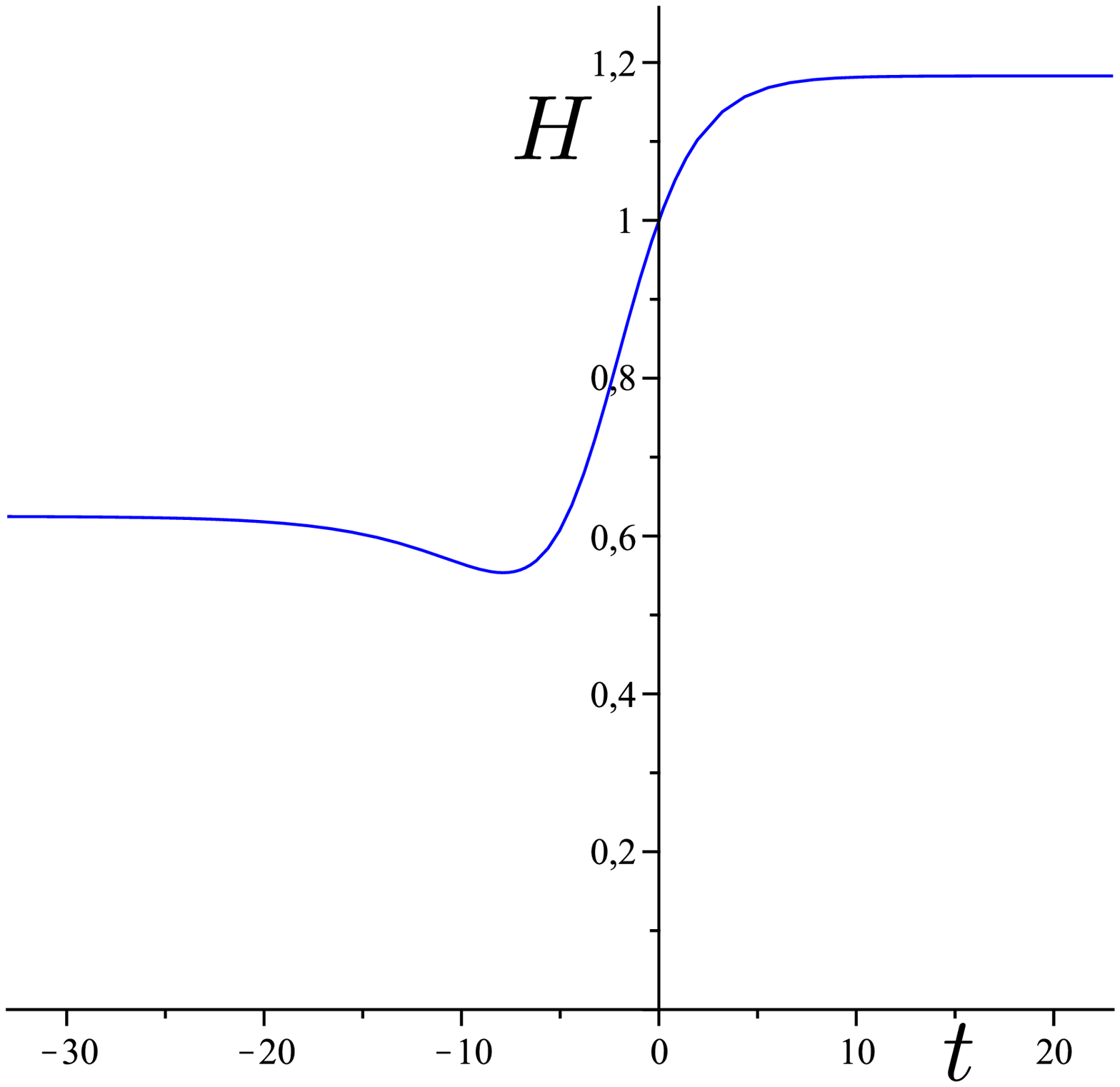} \ \ \ \ \ \ \ \ \ \
\includegraphics[width=45mm]{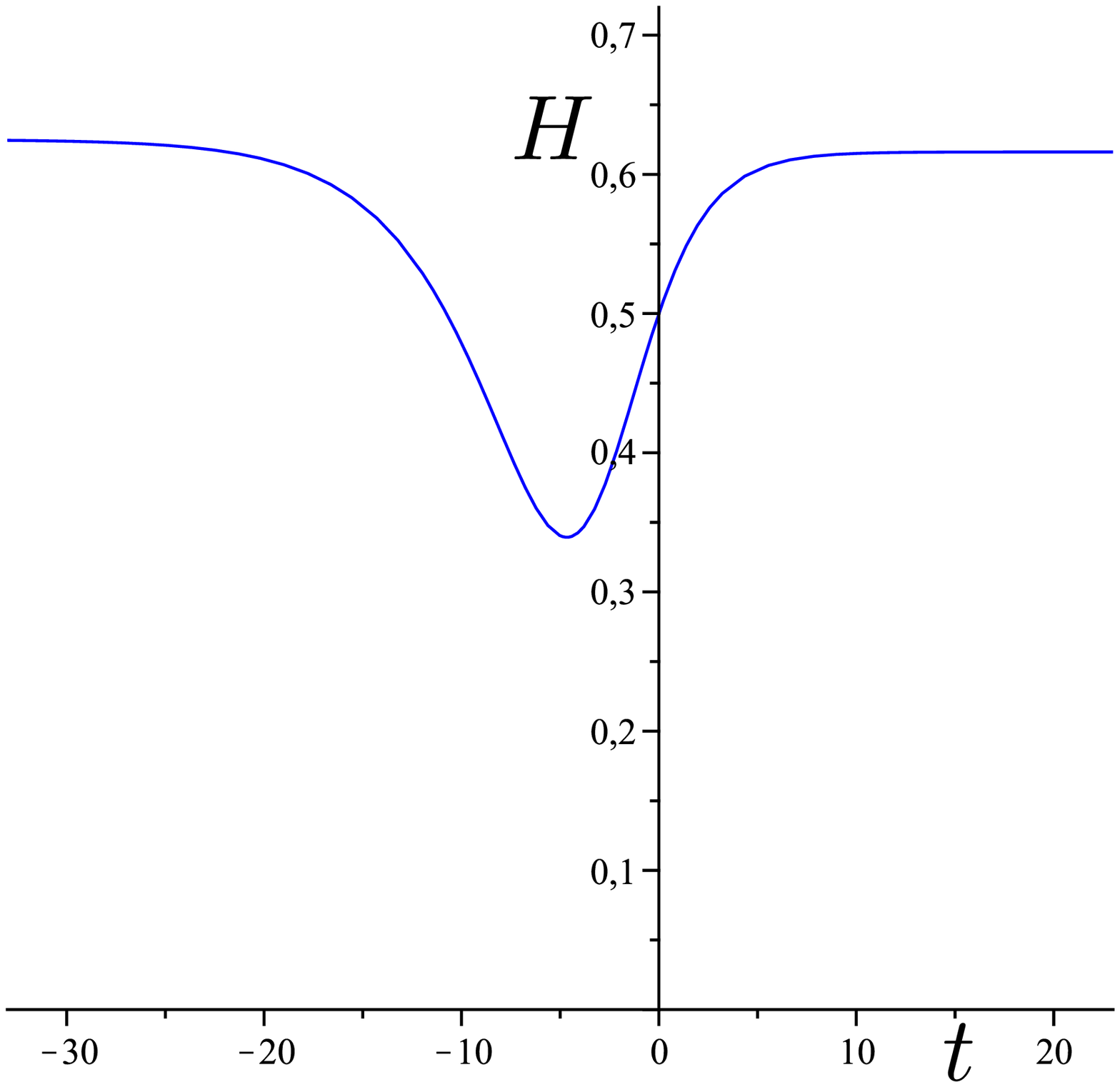} \ \ \ \ \ \ \ \ \ \
\includegraphics[width=45mm]{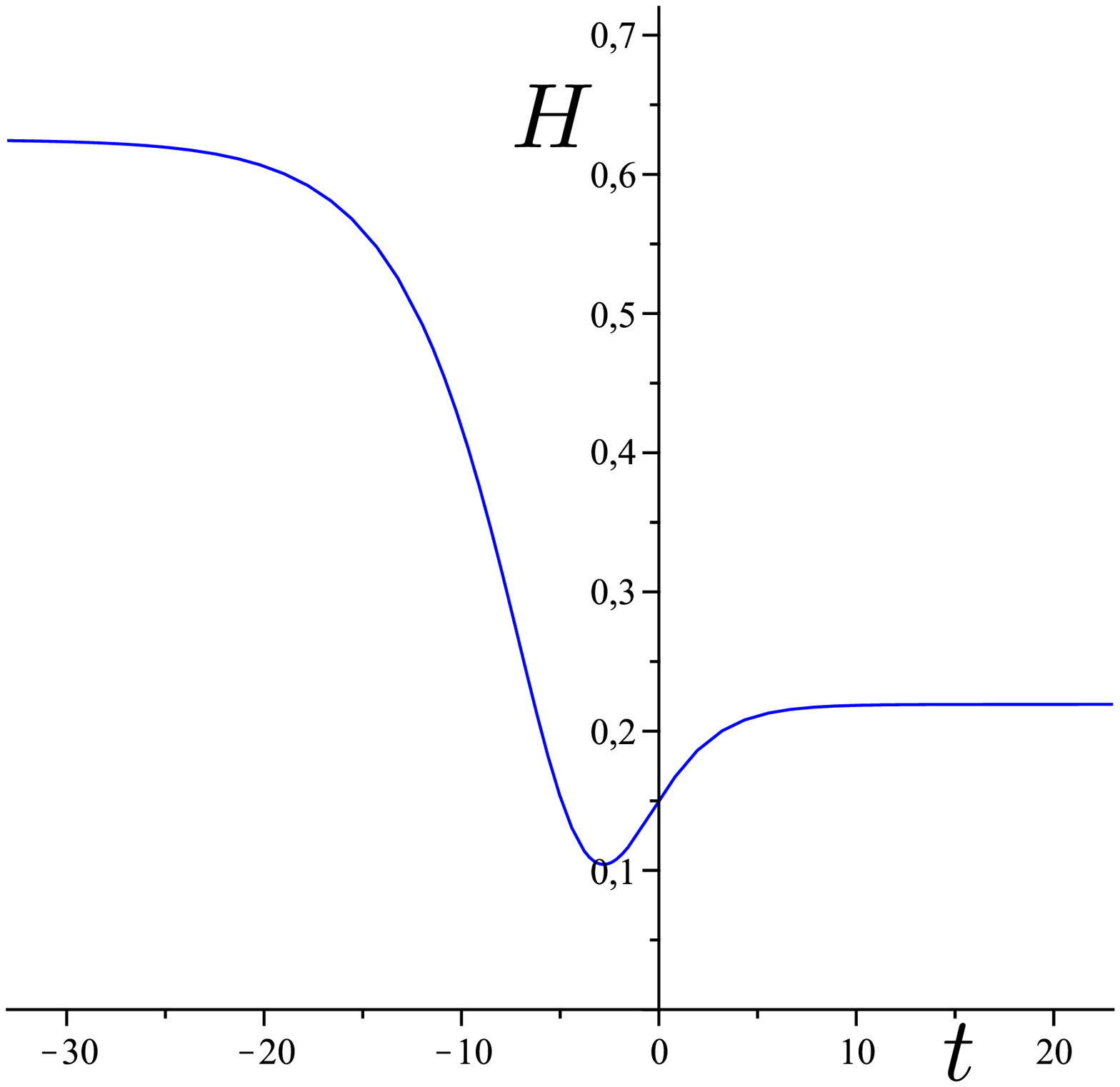}
\caption{Non-monotonic functions $H(t)$, given by (\ref{Y2}) that correspond to $\phi(t)$ plotted in Fig.~\ref{Fbounce}. The values of parameter are $\xi=1$, $C_2=7/2$, and $C_0=5/8$, and $t_0=2\ln(8/9)$. The parameter
  $C_1=-1$ (left), $C_1=-2$ (middle), $C_1=-2.7$ (right).} \label{F2}
\end{figure}
The solution $\phi(t)$ in (\ref{stt}) is associated with different behaviors of the Hubble parameter.
 In Fig.~\ref{F2}, we plot a few non-monotonic $H(t)$ that corresponds to the same $\phi(t)$.
We restrict ourselves to the case $C_0>0$ and $C_2>0$. One can see that the function $H(t)$ is not monotonic if and only~if
\begin{equation}
\label{nonmonotoncond}
\begin{split}
-\frac{2\sqrt{(8\xi+1)C_2C_0}}{\sqrt{4\xi+1}}<C_1<0,  &\quad\mbox{at} \ \phi(t)=\phi_+(t), \\
0<C_1<\frac{2\sqrt{(8\xi+1)C_2C_0}}{\sqrt{4\xi+1}},  &\quad\mbox{at}  \ \phi(t)=\phi_-(t).
\end{split}
\end{equation}
In this case $\dot H=0$ at the point
\begin{equation}
t_1=t_0-\frac{4\xi+1}{4\xi C_0}\ln\left[\frac{\left(4(8\xi+1)C_0C_2-(4\xi+1) C_1^2\right)C_2}{(8\xi+1)C_0C_1^2}\right].
\end{equation}
It is easy to see that the Hubble parameter (\ref{Y2}) is invariant under transformation $C_1 \rightarrow {}-C_1$ and $\phi(t)\rightarrow-\phi(t)$. By this reason $t_1$ is one and the same  both for solutions tend to $\phi_{f_+}$, and for solutions tend to $\phi_{f_-}$.

Solutions described by (\ref{stt}) tend to a fixed point. Indeed, if $\omega>0$, then
\begin{equation}
\label{FP}
\lim_{t\rightarrow\infty}\phi(t)=\phi_{f_\pm}\equiv\pm\frac{\sqrt{(8\xi+1)C_0}}{\sqrt{(4\xi+1)C_2}}\,.
\end{equation}
In the case $\omega<0$, the function $\phi(t)$ tends to zero at late times. In the case $C_0=0$
the function $\phi(t)$ tends to zero as well. Therefore, the Hubble parameter  tends to a constant value at late times for any case. So, the obtained solutions are asymptotically de Sitter ones. We choose $C_0>0$ and $\xi>0$, so $\omega>0$ as well. Thereby, to analyse stability of these solutions we should analyse the stability of de Sitter solutions with $\phi_{f\pm}$.

\section{Stable solutions that tend to a fixed point}

\subsection{The Lyapunov stability}

We consider the stability with respect to homogeneous isotropic perturbations. In other words, we consider system (\ref{IG_SYSTEM}) and see stability with respect to the initial condition.

Let us remind few facts about stability
\cite{Lyapunov,Pontryagin} of solutions for a general system
of the first order autonomic  equations
\begin{equation}
\label{aue}
\dot y_k=F_k(y),\qquad k=1,2,\dots,N.
\end{equation}

By definition a solution (a trajectory)
$y_0(t)$ is attractive (stable) if
\begin{equation}
\|\tilde{y}(t)-y_0(t)\| \rightarrow 0 \quad \mbox{at} \quad
t\rightarrow \infty
\end{equation}
for all solutions $\tilde{y}(t)$ that
start close enough to $y_0(t)$.

We assume that $y(t)$ tends to a fixed point $y_f$.
If all solutions of the dynamical system that start out near a fixed
(equilibrium) point $y_f$,
\begin{equation}
\label{auf} F_k(y_f)=0, \qquad k=1,2,\dots,N
\end{equation}
stay near $y_f$  forever, then $y_f$ is {\it a Lyapunov
stable point}. If all solutions that start out near the  equilibrium
point $y_f$ converge to $y_f$, then the fixed point $y_f$ is {\it an
asymptotically stable} one. Asymptotic stability of fixed point  means
that solutions that start close enough to the equilibrium not only
remain close enough but also eventually converge to the equilibrium.

The Lyapunov theorem~\cite{Lyapunov,Pontryagin} states that to prove
the stability of fixed point $y_f$ of nonlinear system (\ref{aue}) it
is sufficient to prove the stability of this fixed point for the
corresponding linearized system ($y$ is a column):
\begin{equation}
\dot y=A y, \qquad A_{ik}=\frac{\partial F_i(y)}{\partial
y_k}|_{y=y_f}.
\end{equation}
The stability of the linear system means that real parts of all
roots $\lambda_k$ of the characteristic equation
\begin{equation}
\label{equchar}
\det\left(A-\lambda I\right)|_{y=y_f}=0
\end{equation}
are negative. Here $I$ is the identity matrix.

If at least one root of (\ref{equchar}) is positive, then the fixed point is unstable.
The case with pure imaginary eigenvalues of the Jacobian matrix of $F$
at the fixed point requires more specific treatment~\cite{Arnold-Ilush}.

If a solution tends to a stable fixed point, then this solution is stable.
Indeed, if a solution $y(t)$ tends to a fixed point $y_f$, then for any $\varepsilon>0$ there exist $t_1$ such that for all $t>t_1$ $|y(t)-y_f|<\varepsilon/2$.
At the same time if functions $F_k$ are defined and continuous together with their partial derivatives, then solutions of system (\ref{aue}) continuously depend on the initial values~\cite{Pontryagin}. It means that for any finite time $t_1$ and any $\varepsilon>0$ there exist such $\delta$ that $|y(t_1)-\tilde{y}(t_1)|<\varepsilon/2$ for all solutions $\tilde{y}(t)$ such that $|y(t_0)-\tilde{y}(t_0)|<\delta$. We denote the initial moment of time as $t_0$. It means that
that for all solutions $\tilde{y}(t)$ we get $|\tilde{y}(t_1)-y_f|<\varepsilon$, therefore,
$|\tilde{y}(t)-y_f|<\varepsilon $ for all $t>t_1$ and $|y(t)-\tilde{y}(t)|<\varepsilon$ for all $t>t_1$, so $y(t)$ is a stable solution. Therefore, a solution $y_0(t)$ of (\ref{aue}), which tends to the fixed point $y_f$, is attractive if and only if the point $y_f$ is asymptotically stable.

The point $\phi=0$ is a singular point for system (\ref{IG_SYSTEM}), so the above-mentioned arguments are not valid when one considers solutions that tend to this point. We restrict ourselves to consideration of solutions with $\omega>0$ and consider the Lyapunov stability of fixed points, corresponding to $\phi_{f\pm}$.

\subsection{Stability conditions }
To consider the stability of solutions of induced gravity models we denote $y(t)=(\phi(t),\psi(t),H(t))$.
Let $y_f=(\phi_f,\psi_f,H_f)$ is a fixed point. We get $\psi_f=0$. Also from Eqs.~(\ref{e2}) and (\ref{Equ_phi}) we obtain:
\begin{equation}
\label{Vhubble}
V(\phi_f)=3\xi\phi_f^2H_f^2,\qquad V'(\phi_f)=12\xi\phi_fH_f^2,
\end{equation}
consequently,
\begin{equation}
\label{Vcond}
V(\phi_f)=\frac{1}{4}\phi_f V'(\phi_f).
\end{equation}

Let
\begin{equation}
\phi(t)=\phi_f+\varepsilon\phi_1(t),\qquad \psi(t)=\varepsilon\psi_1(t),\qquad H(t)=H_f+\varepsilon H_1(t).
\end{equation}

To first order in
$\varepsilon$ we obtain the following  system of linear equations
\begin{equation}
\begin{split}
\dot{\phi}_1&=\psi_1,\\
\dot{\psi}_1&=\frac{1}{1+6\xi}\left[3\frac{V'(\phi_f)}{\phi_f}-V''(\phi_f)\right]\phi_1-3H_f\psi_1,\\
\dot{H}_1&=\frac{V''(\phi_f)\phi_f-V'(\phi_f)}{(1+6\xi)\phi_f^2}\phi_1+  \frac{4H_f}{(1+6\xi)\phi_f}\psi_1-\frac{24\xi H_f}{1+6\xi}H_1.
\end{split}
\end{equation}

With the result that we get the following matrix
\begin{equation}
A=
\begin{array}{||c c c||}
0 &1 &0\\[7.2mm]
\frac{36\xi H_f^2-V''(\phi_f)}{1+6\xi} & -3H_f &0\\[7.2mm]
\frac{V''(\phi_f)\phi_f-V'(\phi_f)}{(1+6\xi)\phi_f^2}\quad&\quad  \frac{4H_f}{(1+6\xi)\phi_f}\quad
&\quad{}-\frac{24\xi H_f}{1+6\xi}\\
\end{array}
\end{equation}
and roots of Eq.~(\ref{equchar}) are as follows:
\begin{equation}
\begin{split}
\lambda_1&={}-\frac{3}{2}H_f+\frac{\sqrt{9(22\xi+1)H_f^2-4V''(\phi_f)}}{2\sqrt{1+6\xi}},\\
\lambda_2&={}-\frac{3}{2}H_f-\frac{\sqrt{9(22\xi+1)H_f^2-4V''(\phi_f)}}{2\sqrt{1+6\xi}},\\
\lambda_3&={}-\frac{24\xi H_f}{1+6\xi}.
\end{split}
\end{equation}

So, we obtain the conditions on $H_f$ and $V''(\phi_f)$ that are sufficient for the stability
of de Sitter solutions in induced gravity models.
If we assume $\xi>0$, then we get that a fixed point can be stable for $H_f>0$ only.

\section{Stable solutions with a non-monotonic Hubble parameter}

Let us consider the stability conditions for solutions, describing by formulae (\ref{Y2}) and (\ref{stt}).
Using (\ref{Y2}), we get the following conditions on parameters of the potentials:
\begin{equation}
\lambda_3\quad\Leftrightarrow\quad H_f>0\quad\Leftrightarrow\quad\left\{
\begin{array}{ll}
C_1 > -\frac{2(6\xi+1)\sqrt{C_0C_2}}{\sqrt{(4\xi+1)(8\xi+1)}}, &\quad\mbox{for} \quad
\phi_f=\phi_{f_+}\equiv\frac{\sqrt{(8\xi+1)C_0}}{\sqrt{(4\xi+1)C_2}},\\[7.2mm]
C_1 < \frac{2(6\xi+1)\sqrt{C_0C_2}}{\sqrt{(4\xi+1)(8\xi+1)}}, &\quad\mbox{for} \quad
\phi_f=\phi_{f_-}\equiv{}- \frac{\sqrt{(8\xi+1)C_0}}{\sqrt{(4\xi+1)C_2}}.
\end{array}
\right.
\end{equation}

If a solution tends to $\phi_{f_+}$, then
\begin{equation}
\begin{split}
\lambda_{1+}&={}-\frac{3}{2}H_{f_+}+\frac{3\sqrt{C_0(8\xi+1)}}{2\sqrt{C_2(4\xi+1)}}\left(C_1+\frac{2(14\xi+3)\sqrt{C_0 C_2}}{3\sqrt{(8\xi+1)(4\xi+1)}}\right),\\
\lambda_{2+}&={}-\frac{3}{2}H_{f_+}-\frac{3\sqrt{C_0(8\xi+1)}}{2\sqrt{C_2(4\xi+1)}}\left(C_1+\frac{2(14\xi+3)\sqrt{C_0 C_2}}{3\sqrt{(8\xi+1)(4\xi+1)}}\right).
\end{split}
\end{equation}
Analogically, if a solution tends to $\phi_{f-}$, then
\begin{equation}
\begin{split}
\lambda_{1-}&={}-\frac{3}{2}H_{f_-}+\frac{3\sqrt{C_0(8\xi+1)}}{2\sqrt{C_2(4\xi+1)}}\left(C_1-\frac{2(14\xi+3)\sqrt{C_0 C_2}}{3\sqrt{(8\xi+1)(4\xi+1)}}\right),\\
\lambda_{2-}&={}-\frac{3}{2}H_{f_-}-\frac{3\sqrt{C_0(8\xi+1)}}{2\sqrt{C_2(4\xi+1)}}\left(C_1-\frac{2(14\xi+3)\sqrt{C_0 C_2}}{3\sqrt{(8\xi+1)(4\xi+1)}}\right).
\end{split}
\end{equation}
We obtain that all $\lambda_i$ are real. Let us get conditions under that  $\lambda_i$ are negative.
The straightforward calculations show~that
\begin{equation}
\lambda_{1+}={}-\frac{4\xi}{4\xi+1}C_0\,.
\end{equation}
We consider the case $C_0>0$ and $\xi>0$, hence, $\lambda_{1+}<0$. So, the stability of the fixed point $\phi_{f_+}$ depends on sign of
\begin{equation}
\lambda_{2+}={}-3\frac{\sqrt{(8\xi+1)C_0}}{\sqrt{(4\xi+1)C_2}}C_1-\frac{32\xi+6}{4\xi+1}C_0\,.
\end{equation}
We see that
\begin{equation}
\label{STCOND}
    \lambda_{2_+}<0 \qquad \Leftrightarrow\qquad C_1 > {}-\frac{2(16\xi+3)\sqrt{C_0C_2}}{3\sqrt{(8\xi+1)(4\xi+1)}}.
\end{equation}

To explore the stability of the stable point $\phi_{f_-}$, we consider $\lambda_{i_\pm}$ as functions of the parameter $C_1$ and
take notice that $\lambda_{1_-}(C_1)=\lambda_{2_+}(-C_1)$ and $\lambda_{2_-}(C_1)=\lambda_{1_+}(-C_1)$. Consequently, we get
\begin{equation}
\lambda_{1-}=3\frac{\sqrt{(8\xi+1)C_0}}{\sqrt{(4\xi+1)C_2}}C_1-\frac{32\xi+6}{4\xi+1}C_0\,,\qquad
\lambda_{2-}={}-\frac{4\xi}{4\xi+1}C_0\,.
\end{equation}
\begin{equation}
\label{STCOND2}
    \lambda_{1_-}<0 \qquad \Leftrightarrow\qquad C_1 < \frac{2(16\xi+3)\sqrt{C_0C_2}}{3\sqrt{(8\xi+1)(4\xi+1)}}.
\end{equation}

Now we are ready to analyse the stability of the fixed points. Let us start with $\phi_{f_+}$. We consider only the case $C_0>0$, $C_2>0$ and $\xi>0$. We see that $\lambda_{1_+}<0$ and $\lambda_{3_+}<0$ is a more strong restriction on $C_1$ than $\lambda_{2_+}<0$. Therefore, the fixed point $\phi_{f_+}$ is stable at $\lambda_{2_+}<0$. The analogous reasoning gives that $\phi_{f_-}$ is stable at $\lambda_{1_-}<0$.

Now let us analyse the stability of solutions with nonmonotonic Hubble parameter. For such a solution that tends to a stable point $\phi_{f_+}$ condition (\ref{STCOND}) should be satisfied. So, we get a such stable solution at
\begin{equation}
\label{condnonmon}
\begin{split}
 {}-\frac{2(16\xi+3)\sqrt{C_0C_2}}{3\sqrt{(8\xi+1)(4\xi+1)}}< C_1 <0, &\quad \phi(t)=\phi_+(t), \\
0< C_1 < \frac{2(16\xi+3)\sqrt{C_0C_2}}{3\sqrt{(8\xi+1)(4\xi+1)}}, &\quad \phi(t)=\phi_-(t).
\end{split}
\end{equation}
For example, we get that at $\xi=1$, $C_2=7/2$ and $C_0=5/8$, solutions are stable if $C_1>-19\sqrt{7}/18\approx -2.7927$. Therefore, solutions, presented in Fig.~\ref{F2}, are stable, whereas the bounce solution that we plot in Fig.~\ref{Fbounce} is unstable.

\section{Conclusion}

We have analysed the stability of kink-type solutions for the induced gravity
models in the FLRW metric. Using the Lyapunov
theorem we have found sufficient conditions of stability.
 The obtained results allow us to prove that the exact solutions, with non-monotonic behaviors
 of the Hubble parameter, found in~\cite{KTVV2013}, are stable if condition~(\ref{condnonmon}) is satisfied.

Our study  of the stability of isotropic solutions with nonmonotonic behaviors of the Hubble parameter
shows that it is possible to obtain stable solutions with increasing Hubble parameter, in particular, the bounce solutions.

We have analysed the stability of solutions, specifying a form of
fluctuations. It is interesting to know whether these solutions are
stable under the deformation of the FLRW metric to an anisotropic one,
for example, to the Bianchi I metric. Also, it is interesting to check the possibility to get a stable bounce solution
in the induced gravity model. This will be a subject of our future investigations.

\medskip

\noindent {\bf Acknowledgements. }

The authors are grateful to the organizers of the Second Russian-Spanish Congress "Particle and Nuclear Physics at all scales, Astroparticle Physics and Cosmology" for the hospitality and the financial support.
This work is supported in part by the Russian Ministry of Education and Science under grant NSh-3042.2014.2 and by the RFBR grant 14-01-00707.

\end{document}